\documentclass[12pt]{article}

\usepackage{scicite}
\usepackage{times}
\usepackage{braket}
\usepackage{amsmath,amssymb}

\usepackage[dvipdfmx]{graphicx}
\usepackage{dcolumn}
\usepackage{mathptmx,bm}

\usepackage{changes}

\newenvironment{sciabstract}{%
\begin{quote} \bf}
{\end{quote}}

\title{{\bf Title:} Distinct spin and orbital dynamics in Sr$_{2}$RuO$_{4}$}  %No more than 96 characters and spaces%%

\author
{{\bf Authors:} H. Suzuki$^{1,2,3\ast\dagger}$, L. Wang$^{1\ast}$, J. Bertinshaw$^1$, H. U. R. Strand$^{4}$, \\S. K\"{a}ser$^{1,5}$, M. Krautloher$^1$, Z. Yang$^1$, N. Wentzell$^{6}$, O. Parcollet$^{6,7}$, \\F. Jerzembeck$^{8}$, N. Kikugawa$^{9}$, A. P. Mackenzie$^{8}$, A. Georges$^{6,10,11,12}$, \\P. Hansmann$^{1,5,8}$, H. Gretarsson$^{1,13\dagger}$, and B. Keimer$^{1\dagger}$\\
\\[-4mm]
{\bf Affiliations:}\\
\normalsize{$^1$Max-Planck-Institut f\"{u}r Festk\"{o}rperforschung, Heisenbergstra\ss e 1, D-70569 Stuttgart, Germany}\\
\normalsize{$^2$Frontier Research Institute for Interdisciplinary Sciences, Tohoku University,}\\
\normalsize{Sendai, 980-8578, Japan}\\
\normalsize{$^3$Institute of Multidisciplinary Research for Advanced Materials (IMRAM), Tohoku University,}\\
\normalsize{Sendai 980-8577, Japan}\\
\normalsize{$^4$School of Science and Technology, \"{O}rebro University, Fakultetsgatan 1, SE-701 82, \"{O}rebro, Sweden}\\
\normalsize{$^{5}$Department of Physics, Friedrich-Alexander-University (FAU) of Erlangen-N\"urnberg,}\\
\normalsize{91058 Erlangen, Germany}\\
\normalsize{$^6$Center for Computational Quantum Physics, Flatiron Institute, Simons Foundation,}\\
\normalsize{162 5th Avenue, New York 10010, USA}\\
\normalsize{$^7$Universit\'e Paris-Saclay, CNRS, CEA, Institut de physique th\'eorique, 91191, Gif-sur-Yvette, France}\\
\normalsize{$^{8}$Max Planck Institute for Chemical Physics of Solids,}\\
\normalsize{N\"{o}thnitzer Stra\ss e 40, 01187 Dresden, Germany}\\
\normalsize{$^{9}$National Institute for Materials Science, Tsukuba, Ibaraki 305-0003, Japan}\\
\normalsize{$^{10}$Coll\'ege de France, 11 place Marcelin Berthelot, 75005 Paris, France}\\
\normalsize{$^{11}$Centre de Physique Th\'eorique (CPHT), CNRS, Ecole Polytechnique, IP Paris, 91128 Palaiseau, France}\\
\normalsize{$^{12}$Department of Quantum Matter Physics, University of Geneva,}\\
\normalsize{24 Quai Ernest-Ansermet, 1211 Geneva 4, Switzerland}\\
\normalsize{$^{13}$Deutsches Elektronen-Synchrotron DESY,}\\
\normalsize{Notkestra\ss e 85, D-22607 Hamburg, Germany}\\
\\[-4mm]
\normalsize{$^\ast$These authors contributed equally to this work.}\\
\normalsize{$^\dagger$To whom correspondence should be addressed;}\\
 \normalsize{E-mail: hakuto.suzuki@tohoku.ac.jp, hlynur.gretarsson@desy.de, B.Keimer@fkf.mpg.de}
}

\date{}

\begin{document}

\baselineskip24pt

\maketitle 
\clearpage
{\bf Abstract:} %125 words or less
\begin{sciabstract}  
The unconventional superconductor Sr$_2$RuO$_4$ has long served as a benchmark for theories of correlated-electron materials. The determination of the superconducting pairing mechanism requires detailed experimental information on collective bosonic excitations as potential mediators of Cooper pairing. We have used Ru $L_3$-edge resonant inelastic x-ray scattering to obtain comprehensive maps of the electronic excitations of Sr$_2$RuO$_4$ over the entire Brillouin zone. We observe multiple branches of dispersive spin and orbital excitations associated with distinctly different energy scales.
The  spin and orbital dynamical response functions calculated within the dynamical mean-field theory are in excellent agreement with the experimental data. Our results highlight the Hund metal nature of Sr$_{2}$RuO$_{4}$ and provide key information 
for the understanding of its unconventional superconductivity.
 \end{sciabstract}

{\bf One Sentence Summary:} %No more than 125 characters and spaces%%
Resonant inelastic x-ray scattering reveals distinct spin and orbital dynamics in the unconventional superconductor Sr$_{2}$RuO$_{4}$.

\newcounter{Fig1}
\setcounter{Fig1}{1}
\newcounter{Fig2}
\setcounter{Fig2}{2}
\newcounter{Fig3}
\setcounter{Fig3}{3}
\newcounter{hvdep}
\setcounter{hvdep}{1}
\newcounter{sampledep}
\setcounter{sampledep}{2}
\newcounter{DMFT_SOC}
\setcounter{DMFT_SOC}{2}
\newcounter{DMFT}
\setcounter{DMFT}{3}
\newcounter{rpa}
\setcounter{rpa}{4}
\newcounter{bare}
\setcounter{bare}{5}
\newcounter{rixscalc}
\setcounter{rixscalc}{6}

{\bf Main Text:}
For more than a quarter of a century, the layered perovskite Sr$_2$RuO$_4$ has been a cornerstone compound of research on strongly correlated materials. A plethora of experiments performed on exceptionally clean crystals \cite{Mackenzie.A_etal.Rev.-Mod.-Phys.2003} have revealed the emergence of Fermi liquid (FL) quasiparticles at temperatures $T\lesssim$ 30 K \cite{Hussey.N_etal.Phys.-Rev.-B1998} and unconventional superconductivity at $T_c \sim 1.5$ K \cite{Maeno.Y_etal.Nature1994}.
The latter phenomenon attracted considerable experimental and theoretical interest as a potential solid-state realization of spin-triplet superconductivity, analogous to the superfluidity in $^{3}$He \cite{Leggett.A_etal.Rev.-Mod.-Phys.1975,Rice.T_etal.J.-Phys.-Condens.-Matter1995}. 
This notion originated from the lack of reduction of the Knight shift probed by $^{17}$O nuclear magnetic resonance \cite{Ishida.K_etal.Nature1998} and of the magnetic neutron scattering intensity \cite{Duffy.J_etal.Phys.-Rev.-Lett.2000} upon entering the superconducting (SC) state. 
However, recent reinvestigations of these experiments \cite{Pustogow.A_etal.Nature2019,Ishida.K_etal.J.-Phys.-Soc.-Jpn.2020,Petsch.A_etal.Phys.-Rev.-Lett.2020} have revealed a partial reduction of the spin susceptibility, thus reigniting the debate on the order parameter symmetry in the SC state of Sr$_{2}$RuO$_{4}$ \cite{Kivelson.S_etal.npj-Quantum-Mater.2020}.

The orbital degrees of freedom as well as the spin-orbit-coupling (SOC) make the order parameter puzzle particularly intricate, even though the single-particle dispersions are better understood than those of most other metallic or superconducting transition metal oxides.
They lead to multi-sheet Fermi surfaces with mixed orbital and spin character in the normal 
FL state~\cite{Haverkort.M_etal.Phys.-Rev.-Lett.2008,Veenstra.C_etal.Phys.-Rev.-Lett.2014, Zhang.G_etal.Phys.-Rev.-Lett.2016,
Tamai.A_etal.Phys.-Rev.-X2019}, 
which allows a large number of possible pairing symmetries\cite{Kaba.S_etal.Phys.-Rev.-B2019,Ramires.A_etal.Phys.-Rev.-B2019}. 
Dynamical mean-field theory (DMFT) studies\cite{Zhang.G_etal.Phys.-Rev.-Lett.2016,Kim.M_etal.Phys.-Rev.-Lett.2018} 
as well as angle-resolved photoemission spectroscopy experiments\cite{Tamai.A_etal.Phys.-Rev.-X2019} 
have revealed that the effective strength of the SOC is enhanced by electronic correlations. They also render realistic computation of two-particle correlation functions challenging, despite the crucial roles of collective spin and orbital excitations in the unconventional SC pairing.

Inelastic neutron scattering (INS) studies of Sr$_2$RuO$_4$~\cite{Sidis.Y_etal.Phys.-Rev.-Lett.1999,Steffens.P_etal.Phys.-Rev.-Lett.2019,Iida.K_etal.J.-Phys.-Soc.-Jpn.2020,Jenni.K_etal.Phys.-Rev.-B2021}
 have revealed low-energy incommensurate spin fluctuations (ISFs) 
at the in-plane wavevectors ${\bm q}_{\rm{ISF}}=(\pm 0.3, \pm 0.3)$ as well as square-shaped ridge scattering connecting 
them (Fig.\ \arabic{Fig1}a, inset). 
Theoretical studies addressing these results \cite{Boehnke.L_etal.Europhys.-Lett.2018, Strand.H_etal.Phys.-Rev.-B2019, Acharya.S_etal.Commun.-Phys.2019} showed that vertex corrections beyond the perturbative random phase approximation (RPA) are crucial 
to account for the key features of the spin fluctuation spectrum, such as the relative 
suppression of the antiferromagnetic response and, notably,  
the enhancement of the ${\bm q}$-independent (local) contribution to the overall response.  These features highlight the key role played by the Hund's rule coupling in controlling 
the electronic correlations and quasiparticle properties \cite{Strand.H_etal.Phys.-Rev.-B2019}. 
Indeed, DMFT studies \cite{Mravlje.J_etal.Phys.-Rev.-Lett.2011,Kugler.F_etal.Phys.-Rev.-Lett.2020}  
have placed Sr$_2$RuO$_4$ among the broad family of ``Hund metals''
~\cite{Haule.K_etal.New-J.-Phys.2009,Werner.P_etal.Phys.-Rev.-Lett.2008,Medici.L_etal.Phys.-Rev.-Lett.2011,
Hansmann.P_etal.Phys.-Rev.-Lett.2010,Toschi.A_etal.Phys.-Rev.-B2012}, see 
\cite{Georges.A_etal.Annu.-Rev.-Condens.-Matter-Phys.2013} for a review. It was also suggested that the vertex corrections crucially influence the symmetry of the SC pairing \cite{Kaser.S_etal.Phys.-Rev.-B2022}.

One of the distinctive hallmarks of Hund metals is the separation between energy scales 
associated with the onset of coherence of spin and orbital degrees of freedom, respectively, which was predicted theoretically both on a general basis\cite{Stadler.K_etal.Phys.-Rev.-Lett.2015,Horvat.A_etal.Phys.-Rev.-B2016} and 
for the case of Sr$_2$RuO$_4$~\cite{Mravlje.J_etal.Phys.-Rev.-Lett.2016,Kugler.F_etal.Phys.-Rev.-Lett.2020}.  
However, this has only been indirectly inferred from 
the temperature dependence of the Seebeck coefficient \cite{Mravlje.J_etal.Phys.-Rev.-Lett.2016}.
In this report, we take advantage of the unique capability of resonant inelastic x-ray scattering (RIXS) to directly probe the dynamical spin and orbital correlations over a wide range of energy and momentum, and compare the result with state-of-the-art DMFT computations of the spin- and orbital dynamical susceptibilities.

To obtain a comprehensive set of energy-momentum maps of these correlations, 
we have performed RIXS measurements at the Ru $L_3$ edge (2838 eV). 
Figure \arabic{Fig1}A shows the crystal structure of Sr$_{2}$RuO$_{4}$ and the scattering geometry for the RIXS experiment. The incident x-ray photons were $\pi$-polarized, and the scattered photons with both $\sigma$ and $\pi$ polarizations were collected at the scattering angle of 90 degrees. In this geometry, the polarizations of the incident and outgoing photons are always perpendicular, selectively enhancing magnetic responses from the spin and orbital excitations while suppressing the charge response. Given the layered crystal structure of Sr$_{2}$RuO$_{4}$, we express the momentum transfer using the in-plane component ${\bm q}$, which is scanned by changing the sample angle $\theta$. We studied two paths in the reciprocal space, ${\bm q}=(H, 0)$ and $(H, H)$, by fixing the azimuthal angle $\phi$ at 0$^\circ$ and $-45^\circ$, respectively. These paths cross the the ridge and the peak of the low-energy ISFs (inset). The measurements were performed at $T=25$ K, in the FL regime of the normal state.

In Fig.\ \arabic{Fig1}B, we show the Ru $L_3$ RIXS spectra along the two directions. Multiple peak structures are readily identified. The main feature A is composed of multiple peaks which extend up to $\sim$ 1 eV. These peaks are assigned to  spin and orbital excitations within the $t_{2g}$ orbitals. In addition, a weakly-dispersive feature B is identified at $\sim$ 3 eV (blue circles). As this energy corresponds to the splitting of the transitions to the unoccupied 4$d$ $t_{2g}$ and $e_{g}$ orbitals in the Ru $L_3$ x-ray absorption spectrum \cite{SM}, the feature B is readily assigned to the crystal field transitions to the $t_{2g}^{3}e_g^{1}$ electron configurations. We note here that its intensity is maximal close to the ${\bm q}=(0, 0)$ point along the two directions. 

To visualize the characteristics of the RIXS spectra, we show in Fig. \arabic{Fig1}C a colormap of the RIXS intensity. The main feature A is composed of multiple dispersions. Its low-energy tail exhibits downward dispersion toward its local minima at ${\bm q}=(-0.3, 0)$ and $(-0.7, 0)$ along the $(H, 0)$ direction and at ${\bm q}_{\rm{ISF}}=(-0.3, -0.3)$ along the $(H, H)$ direction (white triangles). These ${\bm q}$ vectors are in excellent agreement with those of the ridges and ISFs identified in the previous INS studies \cite{Iida.K_etal.J.-Phys.-Soc.-Jpn.2020,Jenni.K_etal.Phys.-Rev.-B2021}. However, the information from the INS data is limited to the low energy region below $\sim 0.1$ eV, whereas the full access to a large energy window in the present RIXS experiment provides comprehensive information on the ISFs. The colormap also reveals an additional broad dispersive feature C.  It emanates from the top of the feature A at ${\bm q}=(-0.5, 0)$ and $(-0.5, -0.5)$ and merges with the feature B at the $(0, 0)$ point, generating the intensity maximum.  Considering that feature A is composed of spin-orbital excitations within the $t_{2g}$ orbitals and feature B originates from crystal-field transitions, feature C is assigned to orbital fluctuations involving both the $t_{2g}$ and $e_g$ orbitals.

Having identified multiple branches of spin-orbital excitations in Sr$_2$RuO$_4$, we now scrutinize the low-energy excitations within the $t_{2g}$ orbitals. Fig. \arabic{Fig2}A shows an expanded plot of the RIXS spectra below 0.8 eV. The broad global peak maxima (red circles) disperse from 0.2 eV at the zone center ${\bm q}=(0, 0)$, where they are most sharply peaked, to the maximal energy at the zone boundary, ${\bm q}=(0, -0.5)$ and $(-0.5, -0.5)$. We ascribe this dispersion to orbital fluctuations \cite{SM}. Along the $(H, H)$ direction, the low-energy region contains prominent peaks due to the ISFs around ${\bm q}_{\rm{ISF}}=(-0.3, -0.3)$ and subsequent shoulder structures connected to the $(0, 0)$ point (black circles). The quasielastic intensity at the $(0, 0)$ point is significantly weaker than at ${\bm q}_{\rm{ISF}}$, consistent with polarized INS data \cite{Steffens.P_etal.Phys.-Rev.-Lett.2019}. On the other hand, the spin fluctuation intensity is weaker along the $(H, 0)$ direction, except for the small increase of the quasi-elastic intensity of the ridge scattering around $(-0.3, 0)$ and $(-0.7, 0)$ \cite{Iida.K_etal.J.-Phys.-Soc.-Jpn.2020,Jenni.K_etal.Phys.-Rev.-B2021}. In addition, the spectra close to the $(0, 0)$ point contain a broad high-energy tail peaked around $\sim$ 0.5 eV \cite{SM}.  

Figure \arabic{Fig2}B summarizes the ${\bm q}$ dispersions of the observed RIXS features. Along the $(H, 0)$ direction, the orbital fluctuations disperse from $\sim$ 0.2 eV at the $(0, 0)$ point and reach the maximum of $\sim$ 0.5 eV at the $(-0.5, 0)$ point. Along the $(H, H)$ direction, the dispersion is initially steeper and becomes almost flat in the region $H\leq -0.25$ at a higher energy $\sim 0.55$ eV. The spin excitations have a local minimum of 0.06 eV at ${\bm q}_{\rm{ISF}}=(-0.3, -0.3)$ and approach zero energy close to the $(0, 0)$ point. Note here that the spin and orbital fluctuations have distinct energy scales in the entire ${\bm q}$ space without a clear signature of mutual crossing. This observation is of crucial importance in testing the validity of different theoretical approaches. 

To facilitate a direct connection to the INS results, we show in Fig. 2C an expanded colormap of the RIXS intensity around ${\bm q}_{\rm{ISF}}$ and corresponding energy distribution curves with a step size of 0.02 eV. The RIXS intensity around ${\bm q}_{\rm{ISF}}$ shows a conical shape with an isolated intensity maximum at 0.06 eV. The intensity remains centered at ${\bm q}_{\rm{ISF}}$ up to 0.25 eV, consistent with the vertical INS intensity profile at ${\bm q}_{\rm{ISF}}$ observed below $\sim$ 0.06 eV \cite{Iida.K_etal.Phys.-Rev.-B2011}.

We now interpret the low-energy RIXS data in terms of theoretical spin- and orbital susceptibilities, which we computed in DMFT by solving the Bethe-Salpeter equation using the local DMFT particle-hole irreducible vertex within the Ru $4d$-$t_{2g}$ subspace. We employed the same effective model and interaction parameters that have been established in several previous studies \cite{Tamai.A_etal.Phys.-Rev.-X2019,Strand.H_etal.Phys.-Rev.-B2019,Kaser.S_etal.Phys.-Rev.-B2022}.
Theoretical RIXS spectra are constructed by combining the spin and orbital susceptibilities with matrix elements for the RIXS cross section \cite{SM}. 
Figure \arabic{Fig3} shows the comparison of the experimental (panel A) and theoretical DMFT (panel B) RIXS spectra along the previously defined high symmetry momentum paths in the Brillouin zone.  To highlight the importance of the DMFT dynamical vertex corrections, we also show the perturbative RPA spectra (without dynamical vertex) in panel C. It is evident that the DMFT spectra excellently capture the overall dispersion and the distribution of momentum and energy dependent maxima. Specifically, the low-energy intensity is peaked at ${\bm q}_{\rm{ISF}}$ and also extrapolates continuously to the corresponding quasistatic intensity close to zero energy \cite{Strand.H_etal.Phys.-Rev.-B2019}. 
Moreover, the broader maximum emanates from 0.2 eV at ${\bm{q}}=(0, 0)$ and disperses more steeply along the $(H, H)$ direction.
In contrast, the spectral weight distribution in RPA fails to capture the low-energy intensity maximum at ${\bm q}_{\rm{ISF}}$ and yields a spuriously sharp feature that extends from high energy to zero energy around the $\Gamma$ point.
This difference between DMFT and RPA originates from the distinct behavior of the spin- and orbital dynamical responses. We provide the corresponding plots in Figs.\ S\arabic{DMFT_SOC}-S\arabic{bare} of the supplementary materials \cite{SM}. 
While in RPA both the spin- and orbital responses disperse and have spectral weight over the entire energy range, the vertex corrections in DMFT lead to the clear energy separation of spin and orbital contributions predicted for Hund metals. The spin response accounts for almost all the spectral weight at low energies up to $\sim 0.2$ eV, and becomes negligible above this scale.
On the other hand, the orbital response sets in at higher energies $>0.2$ eV and shows pronounced maxima centered around commensurate momenta $\bm{q}=(-0.5, 0)$ and $(-0.5, -0.5)$. 
The RIXS data thus provide direct and quantitative evidence for the spin-orbital separation 
in correlated Hund metals described by DMFT. Further improvement between the experiment and theory could be obtained by a more rigorous treatment of the resonance effect in the RIXS cross section, in particular on the spectral weight maximum of the orbital excitations around $\bm{q}=(-0.5, -0.5)$. 

This result has profound implications for the microscopic description of superconductivity in Sr$_2$RuO$_4$. As primary candidates of bosonic fluctuations mediating the Cooper pairing, the spin and orbital dynamical susceptibilities enter the Eliashberg equations whose self-consistent solution yields the SC order parameter. 
It is therefore crucial for a theory to 
describe the key features of dynamical responses probed by experiments in a wide energy range. Recent theoretical studies suggest that different approximations lead to qualitatively different ground states \cite{Kaser.S_etal.Phys.-Rev.-B2022,Gingras.O_etal.Phys.-Rev.-Lett.2019}. The RIXS data thus point to the critical role of vertex corrections also for the microscopic calculation of the SC order parameter, and invalidate predictions based on the RPA susceptibilities.

In conclusion, we have presented Ru $L_3$ RIXS measurements of the dynamical response functions 
in the unconventional superconductor Sr$_2$RuO$_4$ over a broad range of energy and momentum. 
We have identified several branches of spin and orbital excitations and revealed the separation of 
energy scales associated with these two sets of degrees of freedom, a predicted hallmark of Hund metals 
which had not yet received direct experimental confirmation.  
The measured spectra are in excellent agreement with theoretical calculations based on DMFT including vertex corrections, 
while significant discrepancies with the perturbative RPA approximation are found. 
Our results thus epitomize the power of state-of-the-art many-body theories to yield a detailed, quantitative understanding of complex electronic correlation functions in real materials.
By establishing the properties of key collective modes, they also provide a solid baseline 
for the future identification of the nature and symmetry of SC
order in this prominent model compound.

% Your references go at the end of the main text, and before the
% figures.  For this document we've used BibTeX, the .bib file
% scibib.bib, and the .bst file Science.bst.  The package scicite.sty
% was included to format the reference numbers according to *Science*
% style.

%BibTeX users: After compilation, comment out the following two lines and paste in
% the generated .bbl file. 

\section*{References and Notes}

\renewcommand{\refname}{}
\vspace{-1cm}
\bibliographystyle{Science}

\section*{Acknowledgments:}
We thank A. Damascelli, G. Khaliullin, A. Yaresko, and D. Kukusta for enlightening discussions.
We acknowledge DESY (Hamburg, Germany), a member of the Helmholtz Association HGF, for the provision of experimental facilities. The RIXS experiments were carried out at the beamline P01 of PETRA III at DESY.

\section*{Funding:}
The project was supported by the European Research Council under Advanced Grant No. 669550 (Com4Com) and Grants-in-Aid for Scientific Research from JSPS (KAKENHI) (number 22K13994). H.S. acknowledges financial support from the JSPS Research Fellowship for Research Abroad. H.S. and L.W. acknowledge financial support from the Alexander von Humboldt Foundation. S.K. acknowledges financial support by the DFG project HA7277/3-1. H.U.R.S. acknowledges financial support from the ERC synergy grant (854843-FASTCORR). N.K. is supported by KAKENHI (Grant Nos. 18K04715, 21H01033, and 22K19093), Core-to-Core Program (No. JPJSCCA20170002) from JSPS, and a JST-Mirai Program (Grant No. JPMJMI18A3). The Flatiron Institute is a division of the Simons Foundation.

\section*{Author contributions:}
H.S., L.W., J.B., Z.Y., and H.G. performed the RIXS experiments. M.K., F.J., N.K., and A.P.M. grew the Sr$_{2}$RuO$_{4}$ single crystals. H.S., L.W., and M.K. performed the sample characterizations. H.G. designed the beamline and IRIXS spectrometer. H.S. analyzed the experimental data. H.U.R.S., S.K., N.W., and O.P. developed the computational framework used in the theoretical calculations. S.K, H.U.R.S., A.G., and P.H. carried out the theoretical calculations of the dynamical response functions. H.S. and P.H. constructed the theoretical RIXS intensity from the response functions. H.S., S.K., H.U.R.S., A.G., P.H., and B.K. wrote the manuscript with input from all the co-authors. B.K. initiated and supervised the project.

\section*{Competing interests:}
Authors declare no competing interests.

\section*{Data and materials availability:}
Raw data for all figures in this paper and the supplementary materials are available at desycloud.

%Here you should list the contents of your Supplementary Materials -- below is an example. 
%You should include a list of Supplementary figures, Tables, and any references that appear only in the SM. 
%Note that the reference numbering continues from the main text to the SM.
% In the example below, Refs. 4-10 were cited only in the SM.     

% For your review copy (i.e., the file you initially send in for
% evaluation), you can use the {figure} environment and the
% \includegraphics command to stream your figures into the text, placing
% all figures at the end.  For the final, revised manuscript for
% acceptance and production, however, PostScript or other graphics
% should not be streamed into your compliled file.  Instead, set
% captions as simple paragraphs (with a \noindent tag), setting them
% off from the rest of the text with a \clearpage as shown  below, and
% submit figures as separate files according to the Art Department's
% instructions.

\clearpage

\begin{figure}[htbp] %figure legend should be no longer than 200 words
   \centering
  \includegraphics[width=16cm]{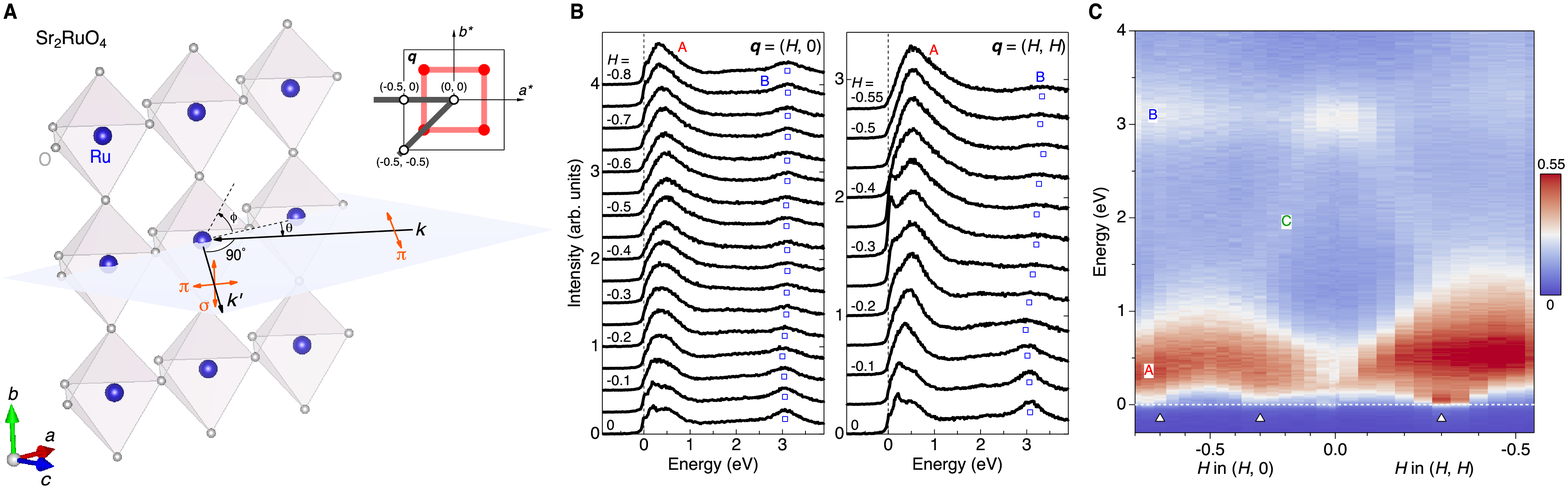}
  %\caption{\raggedright\textbf{Fig. 1. IRIXS of Sr$_{2}$RuO$_{4}$.} {\bf (A)} Crystal structure of Sr$_{2}$RuO$_{4}$ and scattering geometry for the RIXS experiment. Incident x-ray photons with momentum ${\bm k}$ are linearly $\pi$-polarized and the polarization of the scattered photons with momentum ${\bm k^{\prime}}$ is not analyzed. The scattering angle is fixed at 90$^{\circ}$ and the in-plane momentum transfer ${\bm q}$ is scanned by rotating the sample angle $\theta$. The azimuthal angle $\phi$ is used to change the measurement path. The grey lines in the inset indicate the ${\bm q}$ paths in the reciprocal space. Red circles and lines are the schematics of low-energy incommensurate spin fluctuations. {\bf (B)} Ru $L_{3}$ RIXS spectra along the ${\bm q}=(H, 0)$ and $(H, H)$ directions. {\bf (C)} Colormap of RIXS intensity. }
    \label{Sr214rixs}
\end{figure}

\noindent {\bf Fig. 1. IRIXS of Sr$_{2}$RuO$_{4}$.} {\bf (A)} Crystal structure of Sr$_{2}$RuO$_{4}$ and scattering geometry for the RIXS experiment. Incident x-ray photons with momentum ${\bm k}$ are linearly $\pi$-polarized and the polarization of the scattered photons with momentum ${\bm k^{\prime}}$ is not analyzed. The scattering angle is fixed at 90$^{\circ}$ and the in-plane momentum transfer ${\bm q}$ is scanned by rotating the sample angle $\theta$. The azimuthal angle $\phi$ is used to change the measurement paths in the reciprocal space (gray lines, inset). Red circles and lines in the inset are the schematics of the low-energy incommensurate spin fluctuations (ISFs). {\bf (B)} Ru $L_{3}$ RIXS spectra along the ${\bm q}=(H, 0)$ and $(H, H)$ directions. Blue squares indicate the peak positions of crystal field transitions. {\bf (C)} Colormap of RIXS intensity. The positions of ISFs are indicated by open triangles.

\clearpage

\begin{figure}[htbp] %figure legend should be no longer than 200 words
   \centering
  \includegraphics[width=16cm]{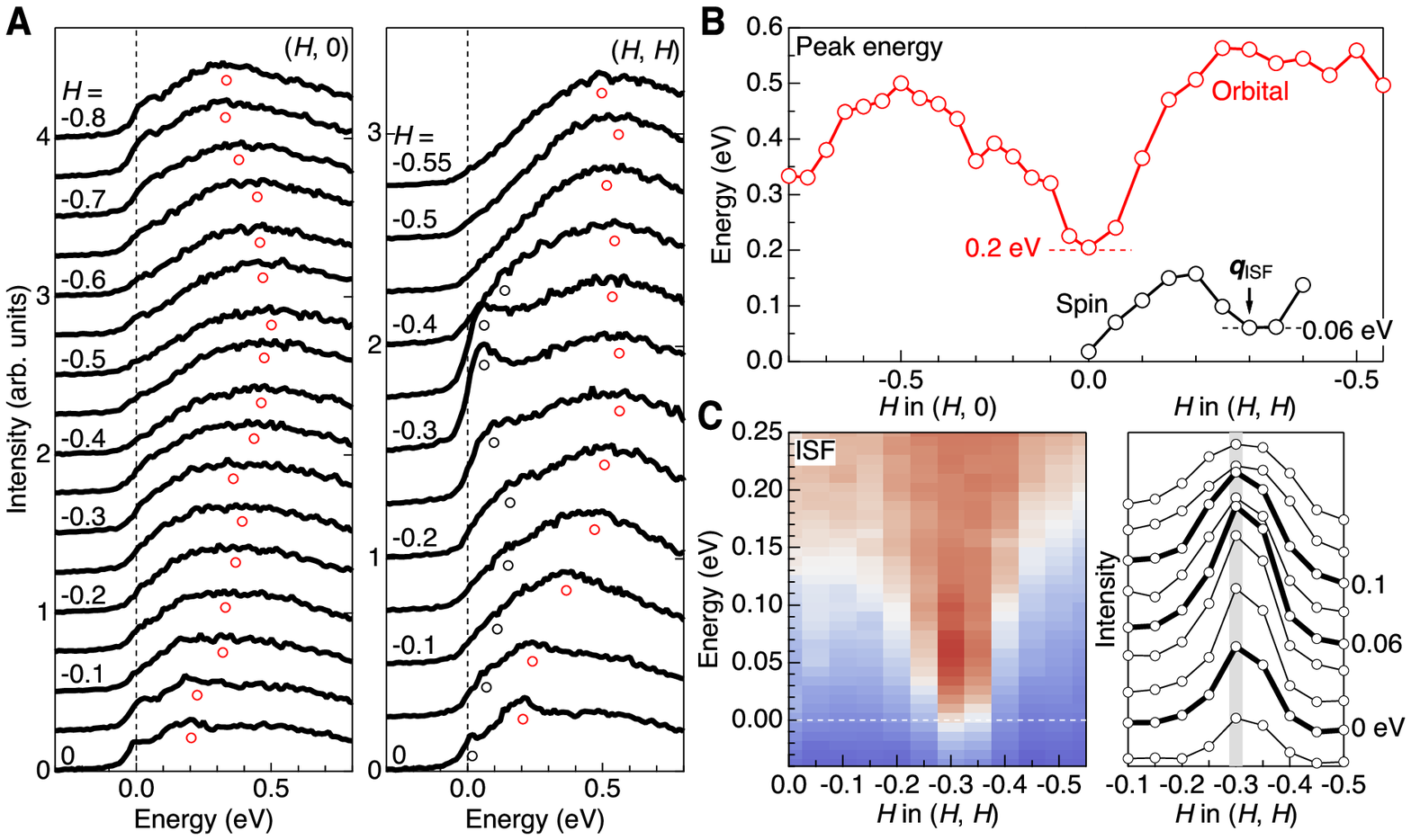}
  %\caption*{\raggedright\textbf{Fig. 2. Incommensurate spin fluctuations.} {\bf (A)} Expanded plots of low-energy RIXS spectra. The incommensurate spin fluctuations are marked with red triangles. {\bf (B)} Second derivative plot of RIXS intensity. The region with largest negative values are associated with incommensurate spin fluctuations (as indicated by triangles) {\bf (C)} Peak energies of dispersive spin and orbital excitations. }
 \label{LowE}
\end{figure}

\noindent {\bf Fig. 2. Spin and orbital fluctuations within the $t_{2g}$ orbitals.} {\bf (A)} Expanded plots of low-energy RIXS spectra. The global peak maxima corresponding to the orbital fluctuations are indicated by red circles. Along the $(H, H)$ direction, the local maxima and shoulder structures from spin fluctuations are indicated with black circles. {\bf (B)} Dispersion relations of the spin and orbital fluctuations as a function of the in-plane momentum transfer. {\bf (C)} Expanded colormap of RIXS intensity around the ISF. The right panel shows momentum distribution curves with a step of 0.02 eV. 

\clearpage

\begin{figure}[htbp] %figure legend should be no longer than 200 words
   \centering
  \includegraphics[width=16cm]{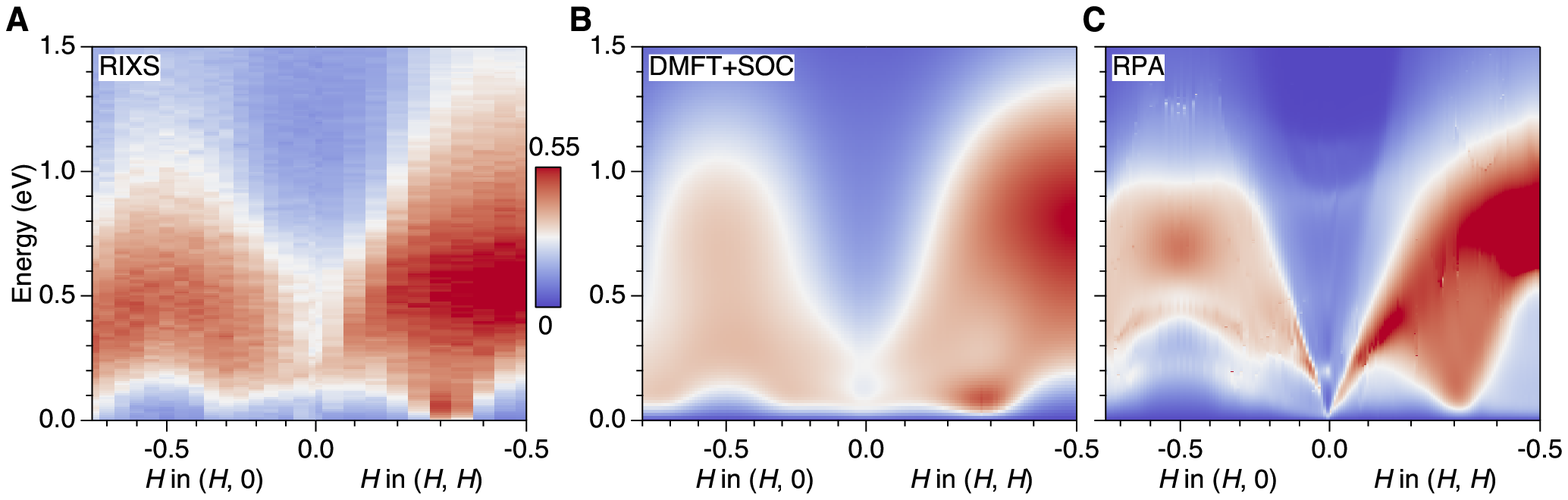}
  %\caption*{\raggedright\textbf{Fig. 3. Modelling of RIXS spectra by theoretical dynamical response functions.} {\bf (A)} Colormap of low-energy RIXS intensity. {\bf (B)} Simulation of RIXS intensity from DMFT+SOC susceptibilities. {\bf (C)} Simulation from RPA susceptibilities.}
 \label{ExpThor}
 \end{figure}

\noindent {\bf Fig. 3. Modelling of RIXS spectra by DMFT+SOC calculations.} {\bf (A)} Expanded colormap of the RIXS intensity within the $t_{2g}^4$ electron configurations. {\bf (B)} Simulation of RIXS intensity based on the spin and orbital susceptibilities calculated by dynamical mean-field theory with spin-orbit coupling (DMFT+SOC). {\bf (C)} Simulation from the susceptibilities calculated with the random phase approximation (RPA).

%%supplementary materials
\clearpage
\setcounter{page}{1}
\setlength{\baselineskip}{18pt}

\begin{center}
\vspace*{2cm}
{\LARGE Supplementary Materials for}\\
\vspace{0.5cm}
{\Large Distinct spin and orbital dynamics in Sr$_{2}$RuO$_{4}$
}\\
\vspace{0.5cm}

% Place the author information here.  Please hand-code the contact
% information and notecalls; do *not* use \footnote commands.  Let the
% author contact information appear immediately below the author names
% as shown.  We would also prefer that you don't change the type-size
% settings shown here.

{H. Suzuki$^{\ast}$, L. Wang, J. Bertinshaw, H. U. R. Strand, S. K\"{a}ser, M. Krautloher, Z. Yang,\\ N. Wentzell, O. Parcollet, F. Jerzembeck, N. Kikugawa, A. P. Mackenzie, A. Georges,\\ P. Hansmann, H. Gretarsson$^{\ast}$, and B. Keimer$^{\ast}$

Correspondence to: hakuto.suzuki@tohoku.ac.jp, hlynur.gretarsson@desy.de, B.Keimer@fkf.mpg.de
}
\end{center}

\noindent{\bf This PDF file includes:}

Materials and Methods

Supplementary Text

Figs. S1 to S6

References

\clearpage
\noindent{\bf Materials and Methods}

\noindent\underline{Single crystals}

The Sr$_{2}$RuO$_{4}$ single crystals with superconducting $T_c\sim 1.5$ K were grown by the floating-zone method \cite{Bobowski.J_etal.Condens.-Matter2019} and pre-aligned using an in-house Laue diffractometer. Sr$_{2}$RuO$_{4}$ has the tetragonal space group $I4/mmm$ with the lattice constants of $a=b=3.903$ and $c=12.901$ \AA. The in-plane momentum transfers are expressed in the reciprocal lattice units (r.l.u.).

\noindent\underline{IRIXS spectrometer}

The RIXS experiments were performed using the intermediate-energy RIXS (IRIXS) spectrometer at the P01 beamline of PETRA III at DESY \cite{Gretarsson.H_etal.J.-Synchrotron-Rad.2020}. The incident x-ray energy was tuned to the Ru $L_{3}$ absorption edge (2838 eV) and incoming photons were monochromatized using an high-resolution monochromator composed of four asymmetrically-cut Si(111) crystals. The polarization of the incident x-ray photons was in the horizontal scattering plane ($\pi$ polarization). The polarizations of the scattered photons were not analyzed. The x-rays were focused to a beam spot of 20 $\times$ 160 $\mu$m$^{2}$ (H $\times$ V). Scattered photons from the sample were collected at the scattering angle of 90$^{\circ}$ (horizontal scattering geometry) using a SiO$_{2}$ (10$\bar{2}$) ($\Delta E$ = 60 meV) diced spherical analyzer with a 1 m arm, equipped with a rectangular [100 (H) $\times$ 36 (V) mm$^{2}$] mask and a CCD camera, both placed in the Rowland geometry. To account for the x-ray self-absorption effect, the RIXS intensity was normalized to the total fluorescent intensity collected with an energy-resolved photon detector placed at the scattering angle of 110$^\circ$. The exact position of the zero energy loss line was determined by measuring non-resonant spectra from silver paint deposited next to the sample. The overall energy resolution of the IRIXS spectrometer at the Ru $L_{3}$-edge, defined as the full width half maximum of the non-resonant spectrum from silver, was $\sim$80 meV. 
%Each measurement was carried out twice to check the reproducibility of the RIXS spectra. 
All the measurements were performed at 25 K (normal state), well above the superconducting $T_{c}$.
\clearpage
\noindent{\bf Supplementary Text}\\
\noindent\underline{Incident energy dependence of RIXS spectra}

Figure S\arabic{hvdep}A shows the Ru $L_3$-edge x-ray absorption spectrum of Sr$_2$RuO$_4$ collected in the total fluorescence yield mode. The data were taken at the sample angle $\theta=30^{\circ}$ at $T=25$ K, with the $\pi$ incident photon polarization. Two features observed at $\sim$ 2838.9 and 2841.3 eV (blue triangles) correspond to the main transitions to the unoccupied 4$d$ $t_{2g}$ and $e_g$ orbitals, respectively. We used 2838 eV (arrow), which is close to the Ru $L_3$-edge absorption threshold, for the RIXS measurements in the main text. Figure S1B shows a colormap of incident-energy dependence of RIXS intensity taken across the Ru $L_3$ edge. The data were taken with a low energy resolution setup ($\Delta E\sim 600$ meV). Close to the $t_{2g}$ resonance (2838.9 eV), the feature below 1 eV is enhanced, whereas close to the $e_{g}$ resonance (2841.3 eV) the feature around 3 eV is enhanced. This observation supports the assignment of the orbital characters of the RIXS features discussed in the main text. Above the $e_g$ resonance, fluorescent-like signals (dashed lines) show up, whose energy loss evolves linearly with the incident energy. These fluorescent-like features originate from nonlocal excitations and bear resemblance to those observed in other perovskite ruthenates Ca$_2$RuO$_4$ \cite{Gretarsson.H_etal.Phys.-Rev.-B2019} and Ca$_3$Ru$_2$O$_7$ \cite{Bertinshaw.J_etal.Phys.-Rev.-B2021}. %Note here that the fluorescent-like signals have vanishing intensity with the incident energy of 2838 eV used in the main text. Instead, the features observed with 2838 eV originate solely from Raman-like features that have constant energy loss as a function of incident energy. This observation supports the interpretation of the observed dispersive RIXS features in terms of spin and orbital dynamical response functions. 
\\

\noindent\underline{Dynamical response functions}

The theoretical calculation of the spin and orbital angular momentum susceptibilities $\chi_{S_\mu S_\mu}$ and $\chi_{L_\mu L_\mu}$ ($\mu = x, y, z$) with Ru $4d-t_{2g}$ symmetry was performed by solving the Bethe-Salpeter equation with the dynamical mean field theory (DMFT) \cite{Georges.A_etal.Rev.-Mod.-Phys.1996} approximation for the particle hole irreducible vertex \cite{Jarrell.M_etal.Phys.-Rev.-Lett.1992,Kunes.J_etal.Phys.-Rev.-B2011,Boehnke.L_etal.Phys.-Rev.-B2011,Lin.N_etal.Phys.-Rev.-Lett.2012} and a bare generalized susceptibility $\chi_0$ containing both DMFT self-energy and correlation enhanced spin-orbit coupling (SOC) corrections \cite{Tamai.A_etal.Phys.-Rev.-X2019, Strand.H_etal.Phys.-Rev.-B2019}. 
The calculations were performed using the two-particle response function toolbox (TPRF) \cite{Strand.H_etal.github} in the imaginary time formalism and analytically continued to real frequency with the maximum entropy algorithm \cite{Jarrell.M_etal.Phys.-Rep.1996} using the \verb|ana_cont| package \cite{Kaufmann.J_etal.2021} and 12 sampled bosonic Matsubara frequencies.
The vertex and self-energy was computed within DMFT without SOC at the temperature $386\,$K, due to technical limitations in the hybridization expansion \cite{Werner.P_etal.Phys.-Rev.-Lett.2006,Werner.P_etal.Phys.-Rev.-B2006,Haule.K_etal.Phys.-Rev.-B2007,Gull.E_etal.Rev.-Mod.-Phys.2011} continuous time quantum Monte Carlo impurity solver \cite{Seth.P_etal.Comput.-Phys.-Commun.2016}.
The effective low energy model was constructed by combining, i) maximally localized Wannier functions using Wannier90 \cite{Marzari.N_etal.Phys.-Rev.-B1997,Mostofi.A_etal.Comput.-Phys.-Commun.2008,Marzari.N_etal.Rev.-Mod.-Phys.2012} and Wien2Wannier \cite{Kunes.J_etal.Comput.-Phys.-Commun.2010} for the three bands crossing the Fermi level with Ru $t_{2g}$ symmetry, and ii) a local Kanamori interaction \cite{Kanamori.J_etal.Prog.-Theor.-Phys.1963} with a Hubbard $U=2.3\,$eV and a Hund's coupling $J=0.4\,$eV \cite{Mravlje.J_etal.Phys.-Rev.-Lett.2011}.
The Wannier construction was performed with an energy window of $[-2.85, 0.75]\,$eV on the band structure from a density functional theory calculation of Sr$_2$RuO$_4$ using the PBE density functional \cite{Perdew.J_etal.Phys.-Rev.-Lett.1996} and Wien2k \cite{wien2kBook} with a $20^3$ k-point grid and the experimental crystal structure (at $100\,$K) \cite{Vogt.T_etal.Phys.-Rev.-B1995}.
The effective model is identical to the one used in \cite{Tamai.A_etal.Phys.-Rev.-X2019, Strand.H_etal.Phys.-Rev.-B2019}.
All calculations were built using the toolbox for interacting quantum systems (TRIQS) \cite{Parcollet.O_etal.Comput.-Phys.-Commun.2015}.

%\textcolor{red}{ToDo: Discuss Figures S\arabic{DMFT_SOC}, S\arabic{DMFT}, S\arabic{rpa}, S\arabic{bare}, S\arabic{rixscalc}.}
%
The resulting DMFT+SOC susceptibility components $\chi_{S_\mu S_\mu}$ and $\chi_{L_\mu L_\mu}$ are shown in figure S\arabic{DMFT_SOC}, and display the energy scale separation between the spin fluctuations at energies $\sim 0.1\,$eV and the orbital fluctuations at energies $\gtrsim 0.2\,$eV with the in-plane response ($xx$ and $yy$) peaking at $\sim 0.75\,$eV while the out-of-plane response peaking at $\sim 0.4\,$eV.
Figure S\arabic{DMFT} shows the effect of neglecting SOC in the bare susceptibility. The magnitude of all components is increased, in particular the incomensurate spin peak at $(H, H) \sim (0.3, 0.3)$, and the spin susceptibility disperses more strongly down in energy at $(0,0)$.
Neglecting the dynamical vertex corrections in DMFT results in the random phase approximation (RPA) with only static interactions and the resulting susceptibilities do not display the spin and orbital angular-momentum energy scale separation, see Fig.\ S\arabic{rpa}.
Static screening was accounted for in the RPA calculation by reducing the local interaction to $U=0.575\,$eV and $J=0.1\,$eV keeping the $J/U$ ratio fixed \cite{Strand.H_etal.Phys.-Rev.-B2019}.
Neglecting interactions all together gives the bare susceptibility, which even lacks the low energy incommensurate spin fluctuations, see Fig.\ S\arabic{bare}.
%
%\ph{ In any case in the supplemental material we need to specify more precisely what is meant by "RPA". Hugo: I agree lets cite Ref. \cite{Strand.H_etal.Phys.-Rev.-B2019} as info on how the RPA was done.}
%
%\textcolor{red}{ToDo: Ensure that Fig.\ S\arabic{rixscalc} is referenced and discussed in the section "Simulation of RIXS intensity".}
\\

\noindent\underline{Simulation of RIXS intensity}

With the theoretical spin and orbital susceptibilities at hand, we have constructed theoretical RIXS intensity in the following way. In general, the RIXS cross section is given by the Kramers-Heisenberg formula \cite{Haverkort.M_etal.Phys.-Rev.-Lett.2010}:
\begin{equation}
\frac{d^2\sigma}{d\Omega d\omega}\propto\sum_{f}\left|\bra{f}T^{\dagger}_{\epsilon_{o}}\frac{1}{\omega_i+E_{i}+i\Gamma-H}T_{\epsilon_{i}}\ket{i}\right|^2\delta(\omega_{i}-\omega_{o}+E_{i}-E_{f}),
\end{equation}
where $H$ is the Hamiltonian and $E_{i}$ ($E_{f}$) is the energy of the initial state $\ket{i}$ (final state $\ket{f}$). $\omega_{i}$ and $\epsilon_{i}$ ($\omega_{o}$ and $\epsilon_{o}$) are the energy and polarization of the incoming (outgoing) photons. $T_{\epsilon}={\bm p}\cdot{\bm A}$ is the optical transition operator, which is expressed as a summation of local transition operators at site $j$: $T_{\epsilon}=\sum_{j}e^{i{\bm k}\cdot{\bm r}_j}T_{j,\epsilon}$. As the core hole is created and annihilated at the same site, the total RIXS transition operator $R^{\epsilon_{i}\epsilon_{o}}=T^{\dagger}_{\epsilon_{o}}\frac{1}{\omega_i+E_{i}+i\Gamma-H}T_{\epsilon_{i}}$ is  expressed as a Fourier transform of local RIXS operators:

\begin{eqnarray}
    R^{\epsilon_{i}\epsilon_{o}}&=&\sum_{j^{\prime},j}e^{i({\bm k}_i\cdot{\bm r}_j-{\bm k}_o\cdot{\bm r}_{j^\prime})}T^{\dagger}_{j^{\prime},\epsilon_{o}}\frac{1}{\omega_i+E_{i}+i\Gamma-H}T_{j,\epsilon_{i}} \nonumber \\ 
    &=&\sum_{j}e^{i{\bm Q}\cdot{\bm r}_j}T^{\dagger}_{j,\epsilon_{o}}\frac{1}{\omega_i+E_{i}+i\Gamma-H}T_{j,\epsilon_{i}}\nonumber \\
    &=&\sum_{j}e^{i{\bm Q}\cdot{\bm r}_j}R^{\epsilon_{i}\epsilon_{o}}_j\\
    &=&R^{\epsilon_{i}\epsilon_{o}}_{\bm Q},
\end{eqnarray}
where ${\bm Q}={\bm k}_i-{\bm k}_o$ is the momentum transfer to the sample. The momentum transfer in the main text is expressed with its in-plane component ${\bm q}$.

As the terms included in $R^{\epsilon_{i}\epsilon_{o}}_j$, we consider only the on-site operators at site $j$ and neglect operators involving the neighboring sites. Furthermore, we employ spherical symmetry approximation at site $j$. Then $R^{\epsilon_{i}\epsilon_{o}}_j$, a bilinear of the components of the polarization vectors $\epsilon_{o}^{\ast}$ and $\epsilon_{i}$, can be decomposed into spherical tensors of different rank:

\begin{equation}
R^{\epsilon_{i}\epsilon_{o}}_j=(\epsilon_{o}^{\ast}\cdot\epsilon_{i})O_j+(\epsilon_{o}^{\ast}\times\epsilon_{i})\cdot {\bm N}_j+\sum_{\mu,\nu=1}^{3}\left[\frac{\epsilon_{o}^{\ast\mu}\epsilon_{i}^{\nu}+\epsilon_{o}^{\ast\nu}\epsilon_{i}^{\mu}}{2}-\frac{1}{3}(\epsilon_{o}^{\ast}\cdot\epsilon_{i})\delta^{\mu\nu}\right]\Xi^{\mu\nu}_{j}, \label{RIXSop}
\end{equation}
where $O_j, {\bm N}_j, \Xi^{\mu\nu}_{j} $ are  the scalar, pseudovector, and quadrupolar operators, respectively. In the present 90$^{\circ}$ scattering geometry (see Fig. \arabic{Fig1}A), $\epsilon_{o}^{\ast}\perp\epsilon_{i}$ holds regardless of the measured ${\bm Q}$. This condition suppresses the scalar (charge) transitions and enhances the pseudovector (magnetic) transitions. We therefore consider only the magnetic channel in the theoretical treatment below. Note, however, that the quadrupolar transitions remain finite.

The orbital degrees freedom in the $t_{2g}^4$ electron configurations of Sr$_2$RuO$_4$, combined with the spin degrees of freedom ($L=1, S=1$), allow a variety of terms in the magnetic channel from a symmetry point of view. This is readily observed in the analytical form of ${\bm N}_j$ under the fast-collision approximation \cite{Kim.B_etal.Phys.-Rev.-B2017}, which includes several combinations of the spin (${\bm S}_j$) and orbital angular momentum (${\bm L}_j$) operators. Here, for simplicity, we approximate ${\bm N}_j$ as a linear combination of spin and orbital angular momentum operators:
\begin{equation}
    {\bm N}_j={\bm S}_j+\alpha {\bm L}_j,
\end{equation}
where $\alpha$ is a fitting parameter accounting for the relative ratio, which varies as a function of the incident x-ray energy (see Fig. S\arabic{hvdep}).

Under this approximation, the theoretical RIXS intensity is expressed as
\begin{eqnarray}
    I({\bm Q},\omega)&=&-\frac{1}{\pi}\mathrm{Im}\left<R^{\epsilon_{i}\epsilon_{o}\dagger}_{\bm Q}R^{\epsilon_{i}\epsilon_{o}}_{\bm Q}\right>\nonumber \\
    &=&-\frac{1}{\pi}\mathrm{Im}\left<R^{\epsilon_{i}\epsilon_{o}}_{-{\bm Q}}R^{\epsilon_{i}\epsilon_{o}}_{\bm Q}\right>, \label{RIXSint}
\end{eqnarray}
with 
\begin{equation}
    R^{\epsilon_{i}\epsilon_{o}}_{\bm Q}=\epsilon_{o}^{\ast}\times\epsilon_{i}\cdot\left({\bm S}_{\bm Q}+\alpha {\bm L}_{\bm Q}\right).
\end{equation}
The polarization vectors $\epsilon_{o}^{\ast}$ and $\epsilon_{i}$ are real-valued in the present case (linear polarizations) and vary with the measured ${\bm Q}$. The RIXS intensity is thus given by a linear combination of the correlation functions of the $S^{\mu}_{\bm Q}$ and $L^{\mu}_{\bm Q}$ $(\mu=x,y,z)$ operators. It is found that the cross terms $\left<S^{\mu}_{-{\bm Q}}L^{\nu}_{\bm Q}\right>$, and the correlators with different indices $\left<S^{\mu}_{-{\bm Q}}S^{\nu}_{\bm Q}\right>$ and $\left<L^{\mu}_{-{\bm Q}}L^{\nu}_{\bm Q}\right>$ $(\mu\neq\nu)$, either vanish by symmetry or are negligibly small. We therefore consider contributions from $\left<S^{\mu}_{-{\bm Q}}S^{\mu}_{\bm Q}\right>$ and $\left<L^{\mu}_{-{\bm Q}}L^{\mu}_{\bm Q}\right>$ to the RIXS intensity [Eq. (\ref{RIXSint})]. To best reproduce the experimental RIXS intensity for each theoretical approximation, the fitting parameter $\alpha$ is set to 3.6 for DMFT+SOC and DMFT susceptibilities, and to 1.3 for RPA and bare susceptibilities. The theoretical RIXS intensities for all these approximations are summarized in Fig. S\arabic{rixscalc}.
\\

\noindent\underline{Origin of 0.5 eV peak around ${\bm q}=(0, 0)$}

%\ph{the temperature difference of experiment and theory should maybe be listed here as well - needs discussion}{\color{red} [Hakuto: Experiment was done at 25 K. Theory at higher temperature, correct?]}

Here we discuss possible origins of the broad 0.5 eV peak observed around ${\bm q}=(0, 0)$, which does not have a corresponding branch in the theoretical RIXS intensity (Fig. \arabic{Fig3}B). We note that the present theoretical model focuses on the electronic degrees of freedom within the $t_{2g}$ orbitals. It is therefore possible that the inclusion of the $e_g$ orbitals to the Hamiltonian and the hybridization between the $t_{2g}$ and $e_g$ orbitals could generate this feature, in addition to the dispersive feature C (Fig. \arabic{Fig1}C). Moreover, a coupling to lattice degrees of freedom could yield additional structures. Finally, we note that the calculation was done at $T\approx 386$ K well above the temperature of the experiment at 25 K (i.e. in the Fermi liquid regime). While such a high calculation temperature would prohibit comparison e.g. to low temperature transport, we do not expect significant temperature effects in the RIXS spectra at finite frequencies.       

Furthermore, the RIXS cross section in the present geometry includes not only the magnetic responses but also the quadrupolar responses, which originate from the last term in eq. (\ref{RIXSop}) and are neglected in the treatment above. The quadrupolar transitions have indeed been observed in a $d^4$ cubic ruthenium Mott insulator K$_2$RuCl$_6$ \cite{Takahashi.H_etal.Phys.-Rev.-Lett.2021}, where spin-orbit transitions from the 
$J=0$ nonmagnetic ground state to the $J=2$ quadrupolar states are located above the main transitions to the $J=1$ magnetic states. This suggests that the quadrupolar operators generally remain active in the Ru $L_3$ RIXS process in the $d^4$ ruthenium compounds with octahedral crystal field environment. Thus, the quadrupolar transitions could contribute to the broad continuum around ${\bm q}=(0, 0)$, which is less pronounced than the main orbital response at 0.2 eV.

\clearpage
\begin{figure}[htbp] %figure legend should be no longer than 200 words
   \centering
  \includegraphics[width=12cm]{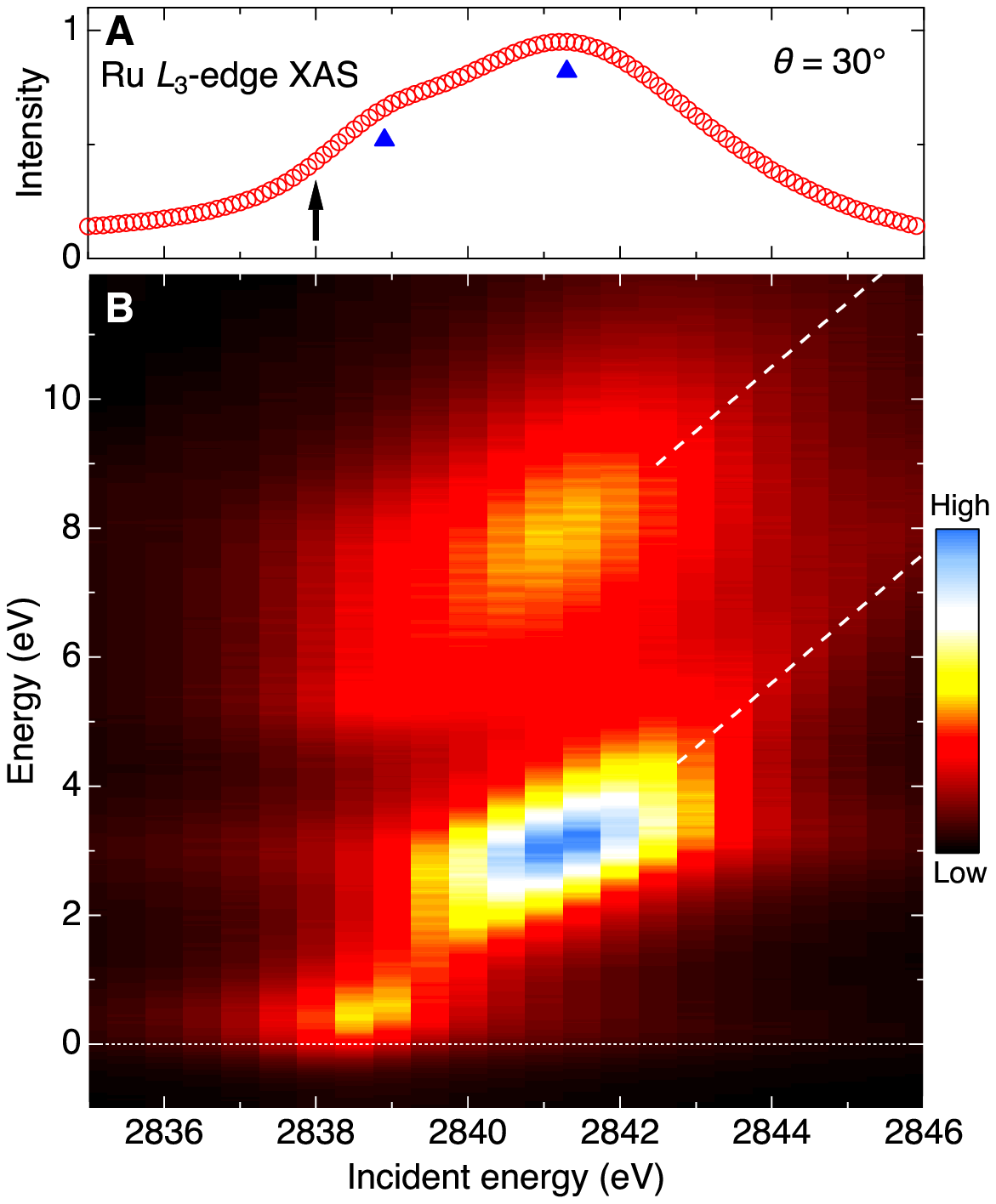}
  %\caption*{\raggedright\textbf{Fig. 3. Modelling of RIXS spectra by theoretical dynamical response functions.} {\bf (A)} Colormap of low-energy RIXS intensity. {\bf (B)} Simulation of RIXS intensity from DMFT+SOC susceptibilities. {\bf (C)} Simulation from RPA susceptibilities.}
\end{figure}

\noindent {\bf Fig. S\arabic{hvdep}.} {\bf (A)} X-ray absorption spectrum of Sr$_2$RuO$_4$ around the Ru $L_3$ edge. The blue triangles indicate the main transitions to the unoccupied Ru 4$d$ $t_{2g}$ and $e_g$ orbitals, respectively. The arrow indicates the incident energy (2838 eV) used for the RIXS measurements in the main text. {\bf (B)} Colormap of the incident-energy dependence of the RIXS spectra across the Ru $L_3$ edge, taken with a low-resolution setup ($\Delta E\sim 600$ meV). The diagonal dotted lines are guides to the eye representing the fluorescent emission. All the data are taken with the incident angle of $\theta=30^{\circ}$ at $T=25$ K.

%\clearpage
%\begin{figure}[htbp] %figure legend should be no longer than 200 words
%   \centering
%  \includegraphics[width=16cm]{RuVacancy.eps}
  %\caption*{\raggedright\textbf{Fig. 3. Modelling of RIXS spectra by theoretical dynamical response functions.} {\bf (A)} Colormap of low-energy RIXS intensity. {\bf (B)} Simulation of RIXS intensity from DMFT+SOC susceptibilities. {\bf (C)} Simulation from RPA susceptibilities.}
%\end{figure}

%\noindent {\bf Fig. S\arabic{sampledep}.} {\bf (A)} Comparison of Raman spectra from Sr$_2$RuO$_4$ crystals with different concentrations of ruthenium vacancies. The triangles indicate additional features that appear in samples with ruthenium vacancies. {\bf (B)} Comparison of Ru $L_3$ RIXS spectra at representative momenta ${\bm q}=(0, 0)$ and $(-0.3, -0.3)$.

\clearpage
\begin{figure}[htbp] %figure legend should be no longer than 200 words
   \centering
  \includegraphics[width=16cm]{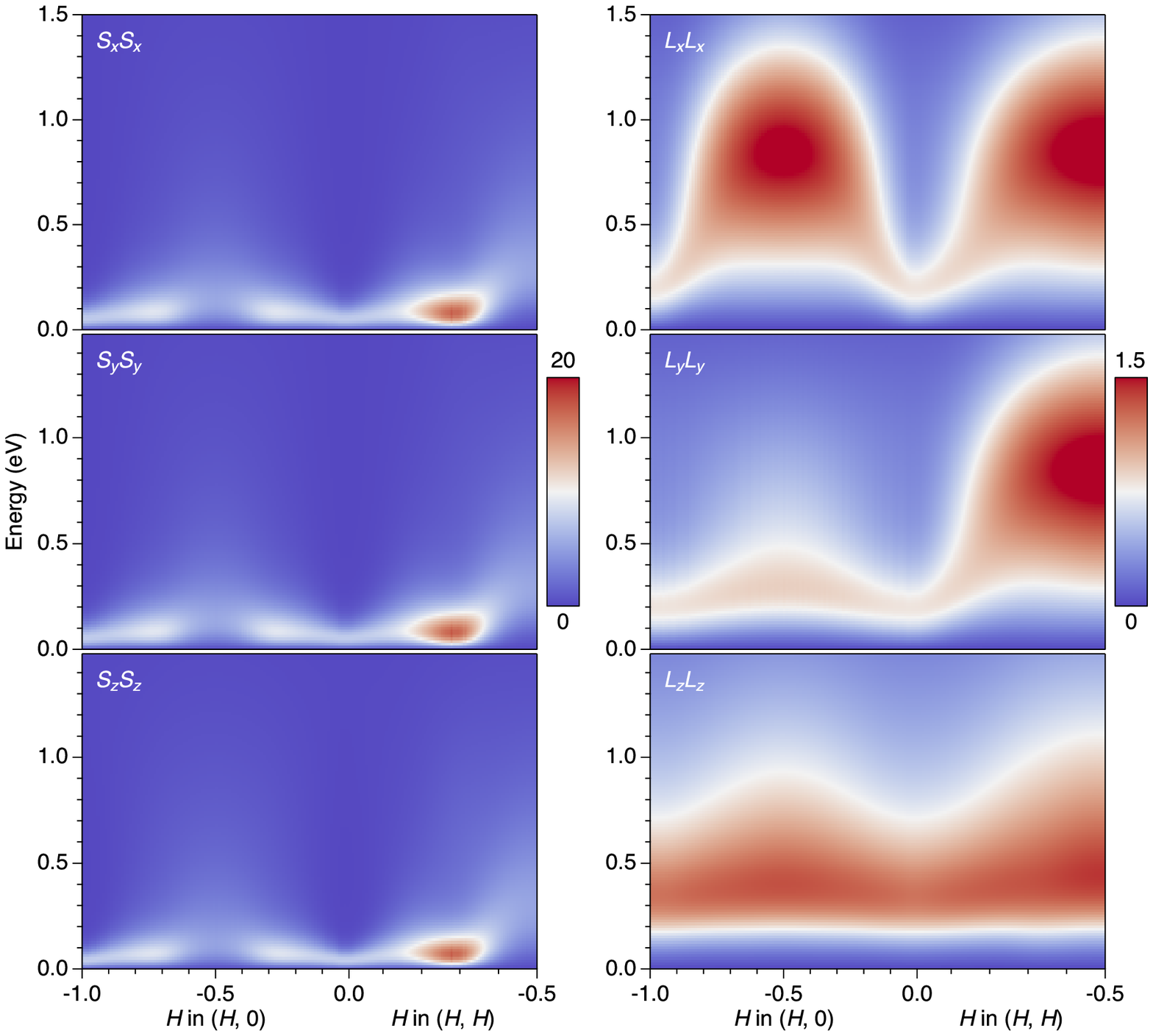}
  %\caption*{\raggedright\textbf{Fig. 3. Modelling of RIXS spectra by theoretical dynamical response functions.} {\bf (A)} Colormap of low-energy RIXS intensity. {\bf (B)} Simulation of RIXS intensity from DMFT+SOC susceptibilities. {\bf (C)} Simulation from RPA susceptibilities.}
\end{figure}

\noindent {\bf Fig. S\arabic{DMFT_SOC}.} Theoretical dynamical mean field theory and spin-orbit coupling (DMFT+SOC) spin and orbital angular-momentum susceptibilities $\chi_{S_\mu S_\mu}$ and $\chi_{L_\mu L_\mu}$ in the plane of the momentum paths $(H,0)$ and $(H, H)$ and energy.

\clearpage
\begin{figure}[htbp] %figure legend should be no longer than 200 words
   \centering
  \includegraphics[width=16cm]{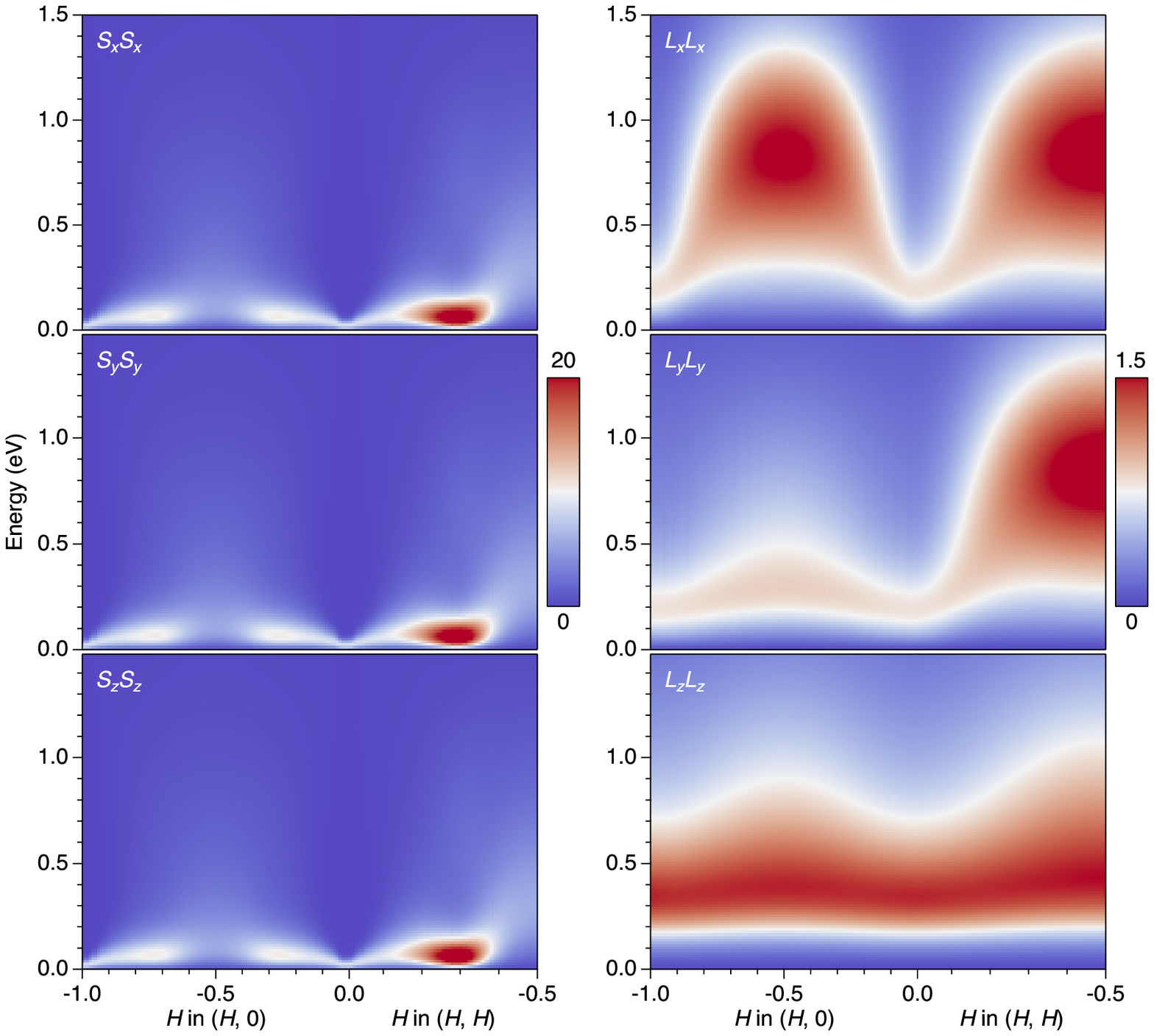}
  %\caption*{\raggedright\textbf{Fig. 3. Modelling of RIXS spectra by theoretical dynamical response functions.} {\bf (A)} Colormap of low-energy RIXS intensity. {\bf (B)} Simulation of RIXS intensity from DMFT+SOC susceptibilities. {\bf (C)} Simulation from RPA susceptibilities.}
\end{figure}

\noindent {\bf Fig. S\arabic{DMFT}.} Theoretical dynamical mean field theory (DMFT) spin and orbital angular-momentum susceptibilities $\chi_{S_\mu S_\mu}$ and $\chi_{L_\mu L_\mu}$ in the plane of the momentum paths $(H,0)$ and $(H, H)$ and energy, showing the result of neglecting spin-orbit coupling (SOC), c.f.\ Fig. S\arabic{DMFT_SOC}.

\clearpage
\begin{figure}[htbp] %figure legend should be no longer than 200 words
   \centering
  \includegraphics[width=16cm]{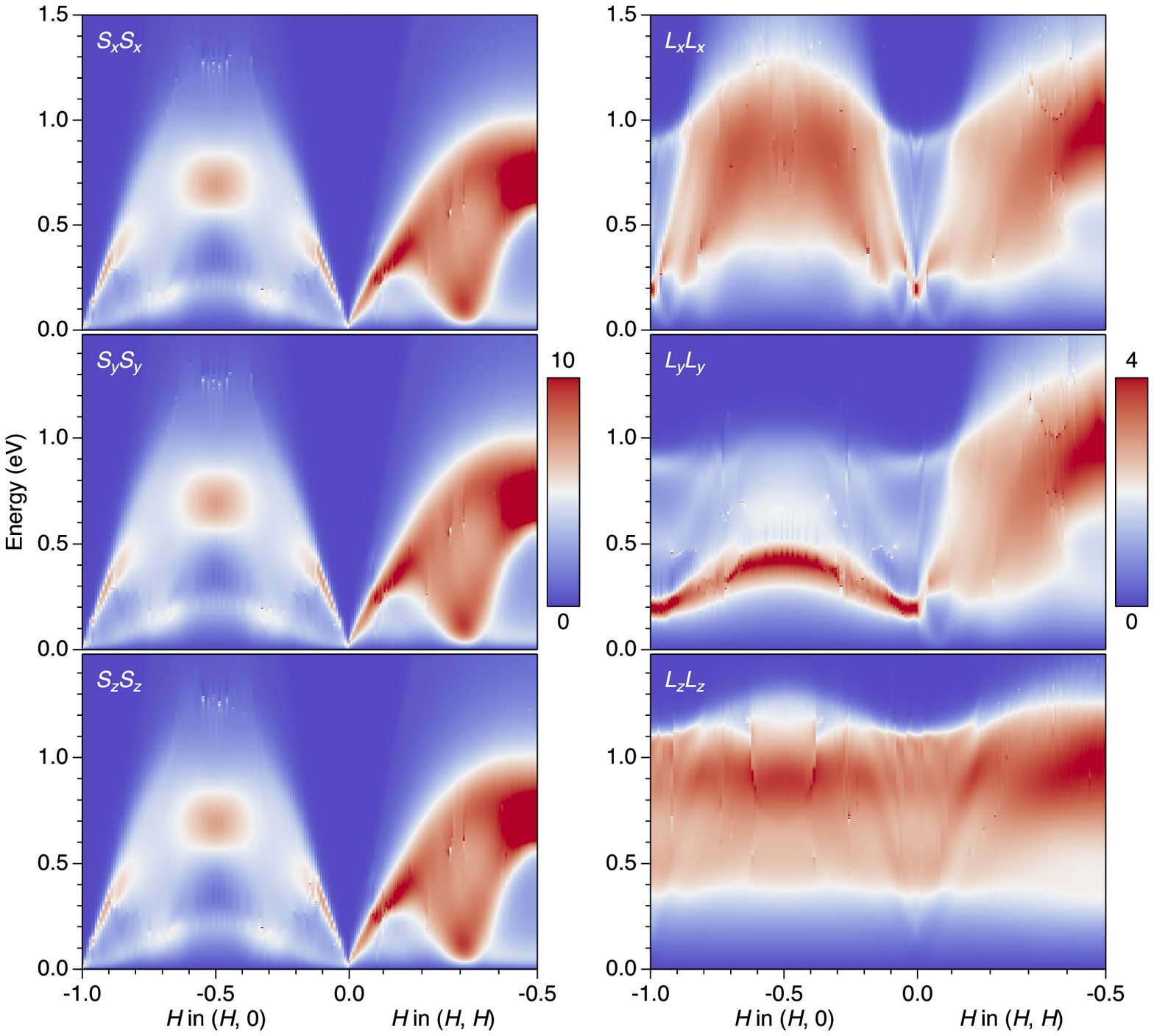}
  %\caption*{\raggedright\textbf{Fig. 3. Modelling of RIXS spectra by theoretical dynamical response functions.} {\bf (A)} Colormap of low-energy RIXS intensity. {\bf (B)} Simulation of RIXS intensity from DMFT+SOC susceptibilities. {\bf (C)} Simulation from RPA susceptibilities.}
\end{figure}

\noindent {\bf Fig. S\arabic{rpa}.} Theoretical random phase approximation (RPA) spin and orbital angular-momentum susceptibilities $\chi_{S_\mu S_\mu}$ and $\chi_{L_\mu L_\mu}$ in the plane of the momentum paths $(H,0)$ and $(H, H)$ and energy, showing the result of neglecting dynamical vertex corrections, c.f.\ Fig. S\arabic{DMFT}.

\clearpage
\begin{figure}[htbp] %figure legend should be no longer than 200 words
   \centering
  \includegraphics[width=16cm]{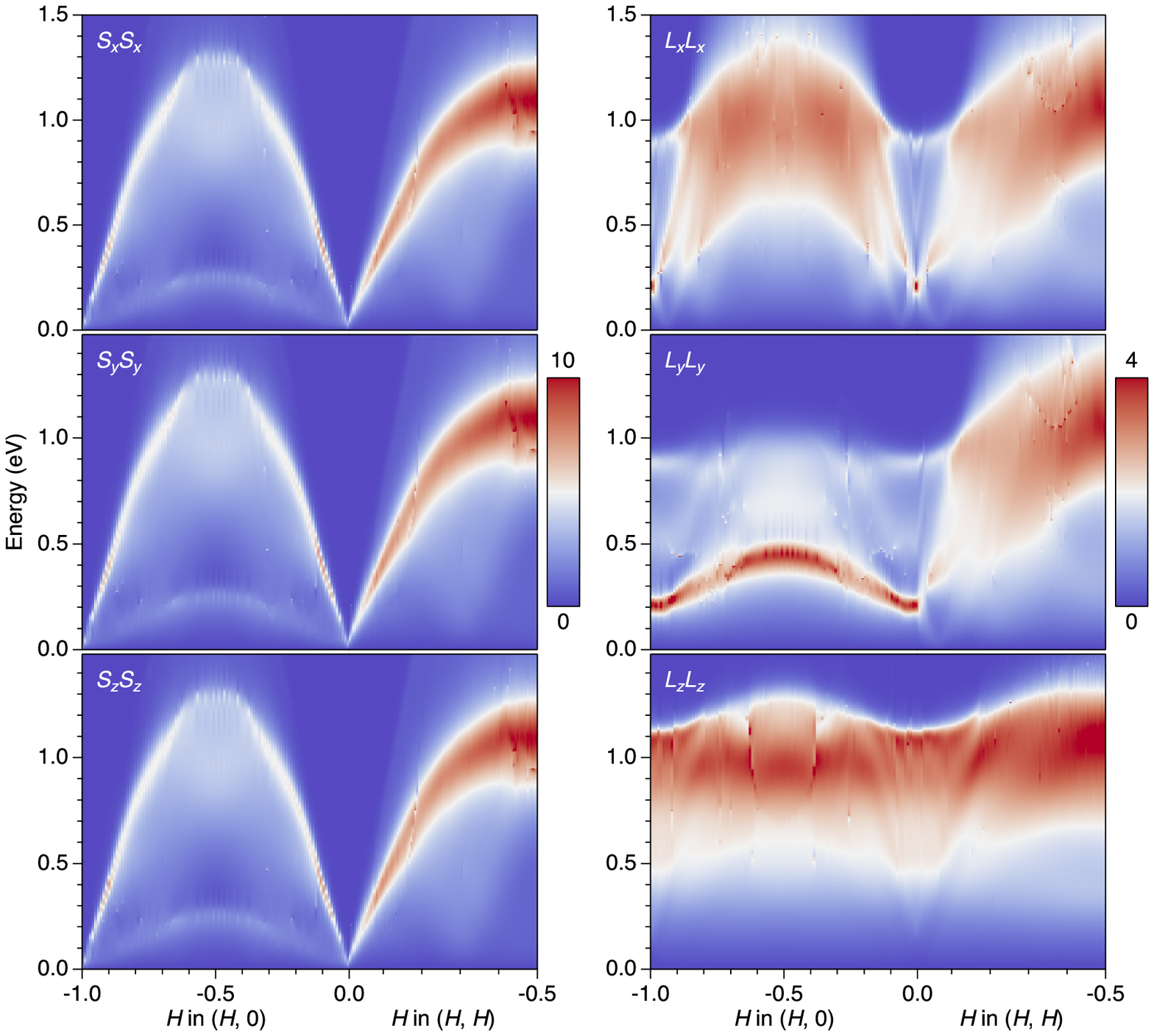}
  %\caption*{\raggedright\textbf{Fig. 3. Modelling of RIXS spectra by theoretical dynamical response functions.} {\bf (A)} Colormap of low-energy RIXS intensity. {\bf (B)} Simulation of RIXS intensity from DMFT+SOC susceptibilities. {\bf (C)} Simulation from RPA susceptibilities.}
\end{figure}

\noindent {\bf Fig. S\arabic{bare}.} Theoretical \emph{bare} spin and orbital angular-momentum susceptibilities $\chi_{S_\mu S_\mu}$ and $\chi_{L_\mu L_\mu}$ in the plane of the momentum paths $(H,0)$ and $(H, H)$ and energy, showing the result of entirely neglecting interactions on the two-particle level, c.f.\ RPA in Fig. S\arabic{rpa} and DMFT in Fig. S\arabic{DMFT}.

\clearpage
\begin{figure}[htbp] %figure legend should be no longer than 200 words
   \centering
  \includegraphics[width=16cm]{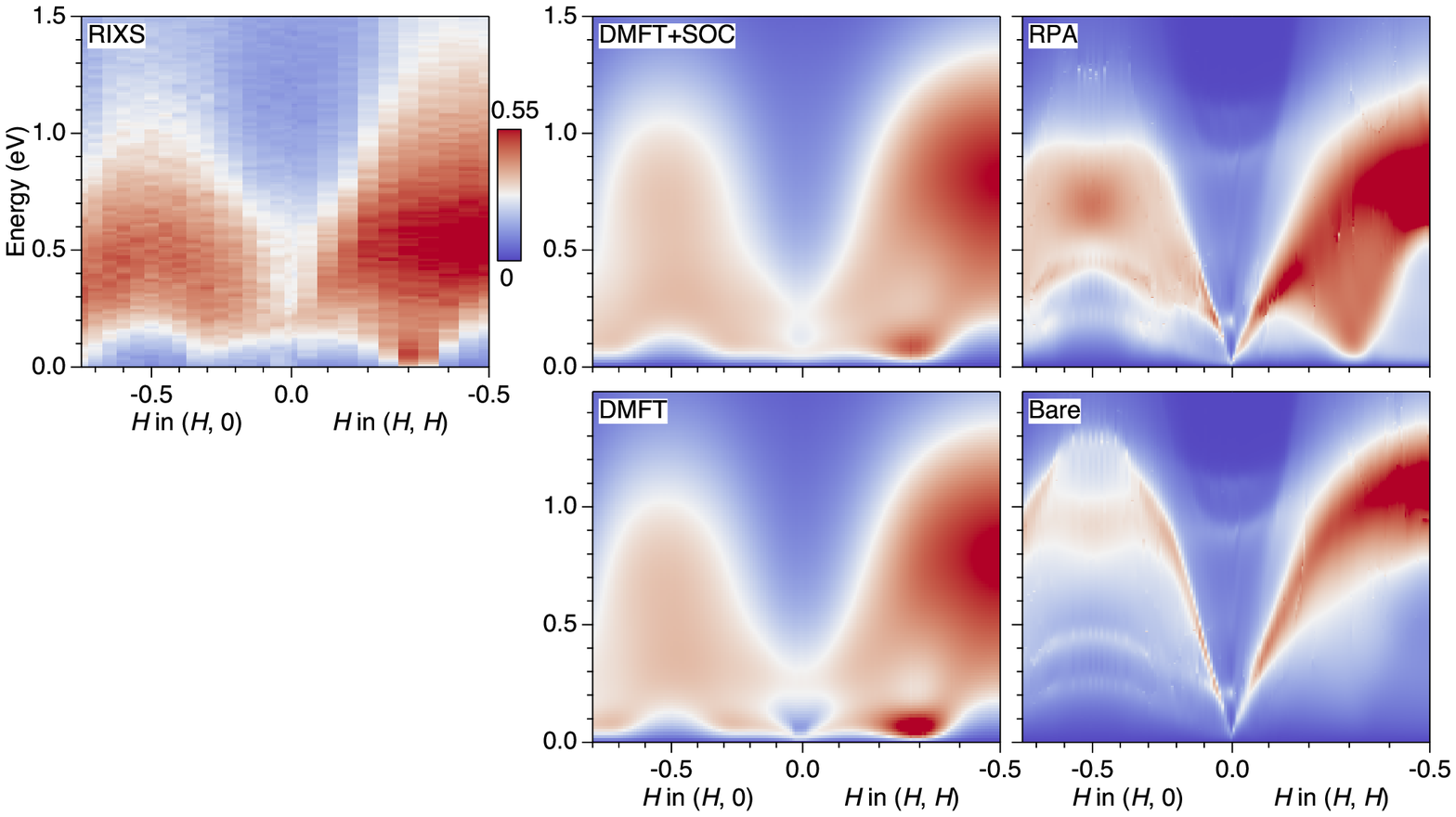}
  %\caption*{\raggedright\textbf{Fig. 3. Modelling of RIXS spectra by theoretical dynamical response functions.} {\bf (A)} Colormap of low-energy RIXS intensity. {\bf (B)} Simulation of RIXS intensity from DMFT+SOC susceptibilities. {\bf (C)} Simulation from RPA susceptibilities.}
\end{figure}

\noindent {\bf Fig. S\arabic{rixscalc}.} Comparison between the
experimenal RIXS spectra {\bf (RIXS)} and the theoretically computed RIXS spectra from DMFT including SOC {\bf (DMFT+SOC)}. Lower levels of theory, such as DMFT without SOC {\bf (DMFT)}, RPA without vertex corrections {\bf (RPA)}, and the bare susceptibility without interactions on the two-particle level {\bf (Bare)}, differ qualitatively from the experimental result.

\clearpage
\noindent{\bf References and Notes}
\vspace{-1.7cm}
\noindent
%\bibliography{scibib}

\end{document}